\documentclass[english]{article}
\usepackage[T1]{fontenc}
\usepackage[latin1]{inputenc}
\usepackage{babel}
\usepackage{graphics}

\makeatletter

\providecommand{\LyX}{L\kern-.1667em\lower.25em\hbox{Y}\kern-.125emX\@}

 \newcommand{\lyxaddress}[1]{
   \par {\raggedright #1 
   \vspace{1.4em}
   \noindent\par}
 }

\makeatother
\begin{document}

\title{Dominant Contributions \\
to Lateral Distribution Functions in \\
Ultra-High Energy Cosmic Ray Air Showers}

\author{Hans-Joachim Drescher and Glennys R. Farrar}

\maketitle

\lyxaddress{\centering Center for Cosmology and Particle Physics\\
Department of Physics, New York University\\
4 Washington Place, New York, NY 10003}

\begin{abstract}
In hadron induced air showers of highest energies (E>\( 10^{18} \)
eV), the lateral distribution functions of electrons and muons are
a superposition of many separate electromagnetic sub-showers, initiated
by meson decay at different altitudes and energies. The lateral distribution
function is the primary tool for reconstructing the energy of the
primary in a UHE cosmic ray shower, so understanding it in detail
is a prerequisite for having confidence in the energy determination.
We analyze in this paper the dominant contributions to the ground
level lateral distribution functions, as a function of the altitude
and energy at which the sub-showers are initiated. Far from the core,
the dominant contribution to the density of electrons comes from sub-showers
initiated at low altitudes and low energies (E<100 GeV). The dominant
sub-showers are initiated at large radial distance from the core and
at a large angle with respect to the main shower axis. This demonstrates
the need for careful treatment of low energy hadron physics modeling
even for ultrahigh energy primaries.
\end{abstract}

\section{Introduction}

There are basically two methods to measure air showers induced by
cosmic rays with an energy greater than \( 10^{18} \) eV. One is
to measure the fluorescence light caused by charged particles passing
through the atmosphere; the other is to sample the lateral distribution
of charged particles on the ground. Theoretical predictions for both
the longitudinal and lateral distributions of secondaries are produced
by Monte Carlo simulation of the showers. Comparing theory and observation
permits properties of the primary cosmic ray, such as its energy and
composition, to be inferred. 

In this paper we study lateral distribution functions of electrons
and muons. In particular we want to determine which altitudes and
which regimes of energy of the hadronic part of an air shower are
of primary importance in producing various components of the ultimate
ground level lateral distribution. This is a step toward the larger
goal of elucidating the sources of systematic uncertainties in theoretical
predictions for shower properties.

\section{Air showers}

\label{sec:air_showers}Atmospheric showers can be categorized into
three elements, the hadronic part, the electromagnetic part and the
muonic part. Basically, the hadronic part feeds the other two components,
photons being mainly produced by \( \pi ^{0}/\eta  \) decay and muons
by the decay of charged pions and kaons. Once a photon is produced
it initiates an electromagnetic sub-shower which evolves essentially
independently of the rest of the shower. (With some low probability,
a photon interacts hadronically in a photo-nuclear reaction, re-initiating
a hadronic sub-cascade; this is neglected here.) Muons, once created,
interact very little. For them the most important process is energy
loss via ionization, and occasionally pair production or bremsstrahlung
in the nuclear field of air molecules. 

Let us call the photons and muons which are directly produced by hadronic
decays, \emph{initiators}, because they initiate a sub-shower of their
kind. The questions we address here are: where are these initiators
produced, what are their energies, and what are the energies of the
hadronic reaction in which they were themselves created. This is crucial
to understanding lateral distribution functions and especially what
governs the particle densities at large distances from the shower
core.

For electromagnetic sub-showers there are two classes of initiators:
high energy photons produced at high altitudes and low energy photons
produced at low altitudes because high altitude initiators must have
sufficient energy for their secondaries to reach ground level before
being absorbed and sub-showers initiated at low energy must be created
at low altitude, to avoid being absorbed before reaching the ground.
However the the relative importance of these classes of initiators,
as a function of the radial distance of the observed particle from
the sub-shower, and the dominant sources of lateral spreading and
arrival-time delay, has not been systematically studied until now. 

For muons the situation is simpler than for electromagnetic showers.
High energy charged pions rarely decay due to their huge Lorentz factor,
so only low energy components of the hadronic shower contribute significantly
to muon production. Moreover Lorentz beaming means that lower energy
particles have typically larger angles with respect to the shower
axis, and therefore contribute more to the muon density at larger
distance from the core, than do higher energy particles. For a given
initiator energy and angle, simple geometry implies that a muon can
propagate to larger distances from the core if it is produced at higher
altitudes. Muons, in contrast to electrons and photons propagate without
much absorption to the ground level.

The air showers analyzed in this paper are simulated with a traditional
Monte Carlo method, whose code will be released soon. The basic features
of this model are the high energy hadronic model QGSJET \cite{QSJET},
the low energy hadronic model GHEISHA \cite{GHEISHA} and the electromagnetic
shower model EGS4 \cite{EGS4}. The model has also some new simulation
techniques which speed up the computation of air showers considerably,
but these are irrelevant for the present analysis. Therefore, the
pure physics content (hadronic and electromagnetic modeling) is identical
to the most commonly used option in the CORSIKA package \cite{CORSIKA}.
We have directly confirmed the agreement of the results presented
here with CORSIKA simulations.

\section{Initiators }

In the following we analyze the contributions to the lateral distribution
functions of \( 5\times 10^{19} \) eV proton-induced vertical showers
at a ground level altitude of 900m, averaging over 50 showers. The
kinetic energy cutoffs for registering hadrons, muons, electrons/positrons
and photons have been chosen to 0.05 GeV, 0.05 GeV, 1 MeV and 1 MeV,
respectively.

\subsection{Ground electrons}

\begin{figure}
{\centering \resizebox*{!}{0.3\textheight}{\includegraphics{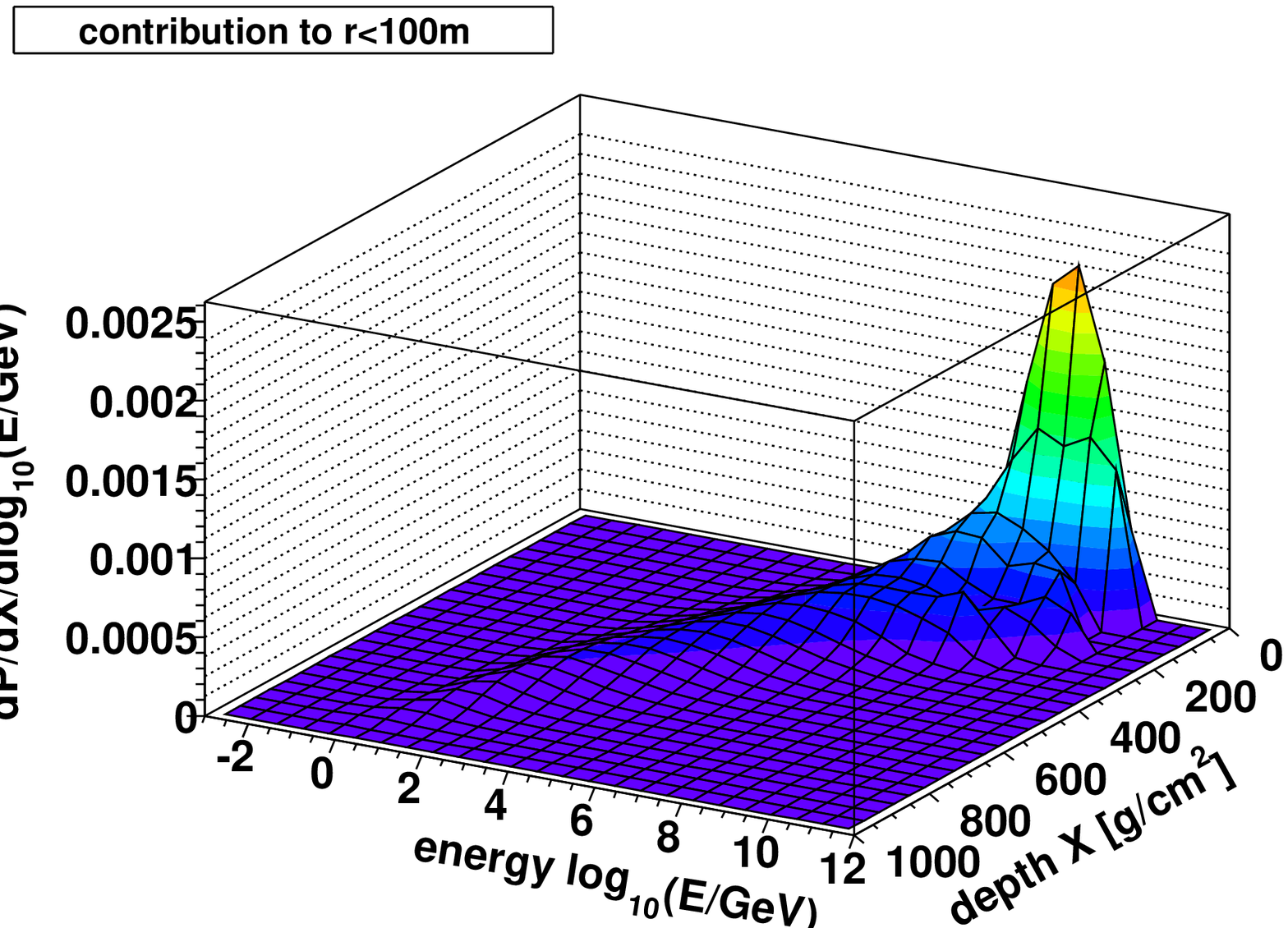}} \par}

{\centering \resizebox*{!}{0.3\textheight}{\includegraphics{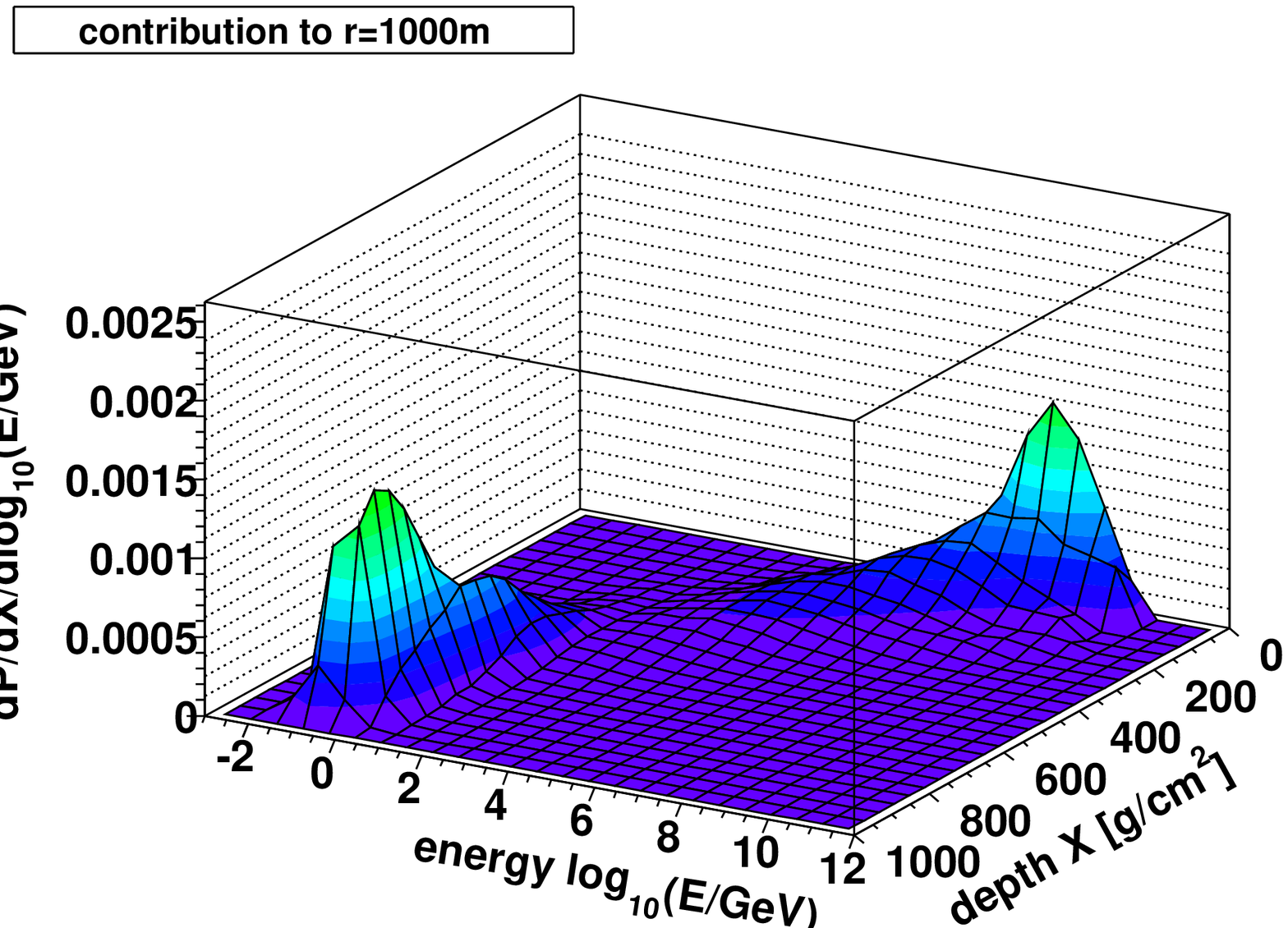}} \par}

{\centering \resizebox*{!}{0.3\textheight}{\includegraphics{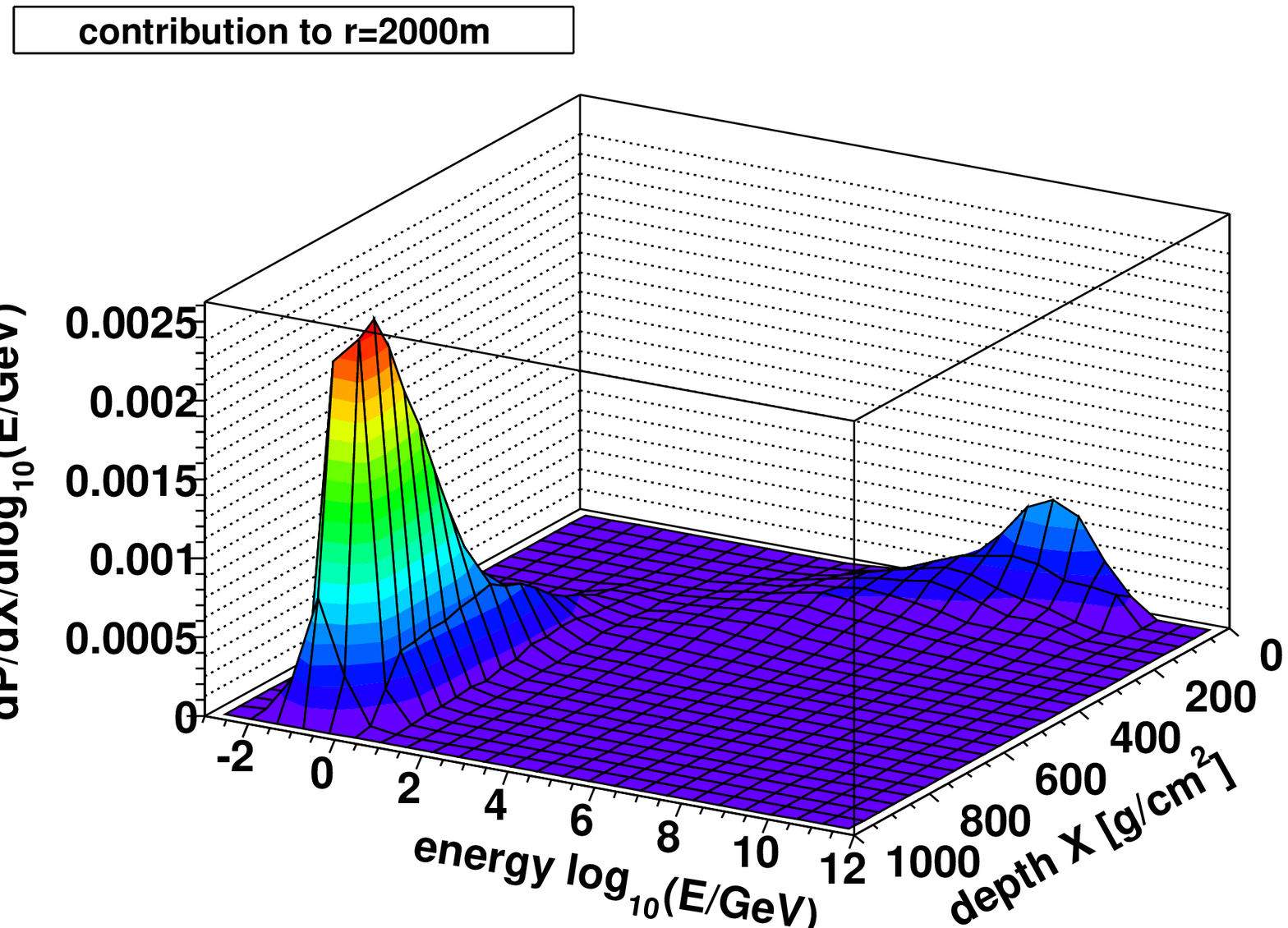}} \par}

\caption{\label{fig:2dplot}The relative contribution to electron densities
at three distances from the core r<100, r=1000, and r=2000m, as a
function of energy and production depth of the initiator of the electromagnetic
sub-shower. Two distinct regions are visible, one at low energy and
low altitude, and another at high altitude and high energy. }
\end{figure}

For electrons at ground level, the results are best summarized in
Fig. \ref{fig:2dplot}, which shows the relative contribution at three
distances from the core as a function of the electromagnetic initiator's
energy and production depth. This is just the probability that an
electromagnetic sub-shower is initiated at a given altitude and energy.
The altitude is plotted in units of vertical atmospheric depth, defined
as \( X(h)=\int _{h}^{\infty }\rho (h')dh' \), \( h \) being the
altitude and \( \rho (h) \) the density of air, according to a standard
USGS atmosphere. The results are obtained as follows: for each electron
arriving at a certain distance from the shower core we trace back
the history of this particle, generation by generation, and record
the energy and depth of production of the initiator (the last particle
of the sub-shower which was not a hadron: in most cases, a photon
from \( \pi ^{0}/\eta  \) decay).

One recognizes two distinct regions of initiator energy and depth
of production which play the dominant role in producing electrons,
depending on the distance from the core. Closer to the core, the initiators
come predominantly from high altitudes and have high energies, while
far from the core the initiators are seen to be primarily produced
at low altitude with low energy. 

\begin{figure}
{\centering \resizebox*{10cm}{!}{\includegraphics{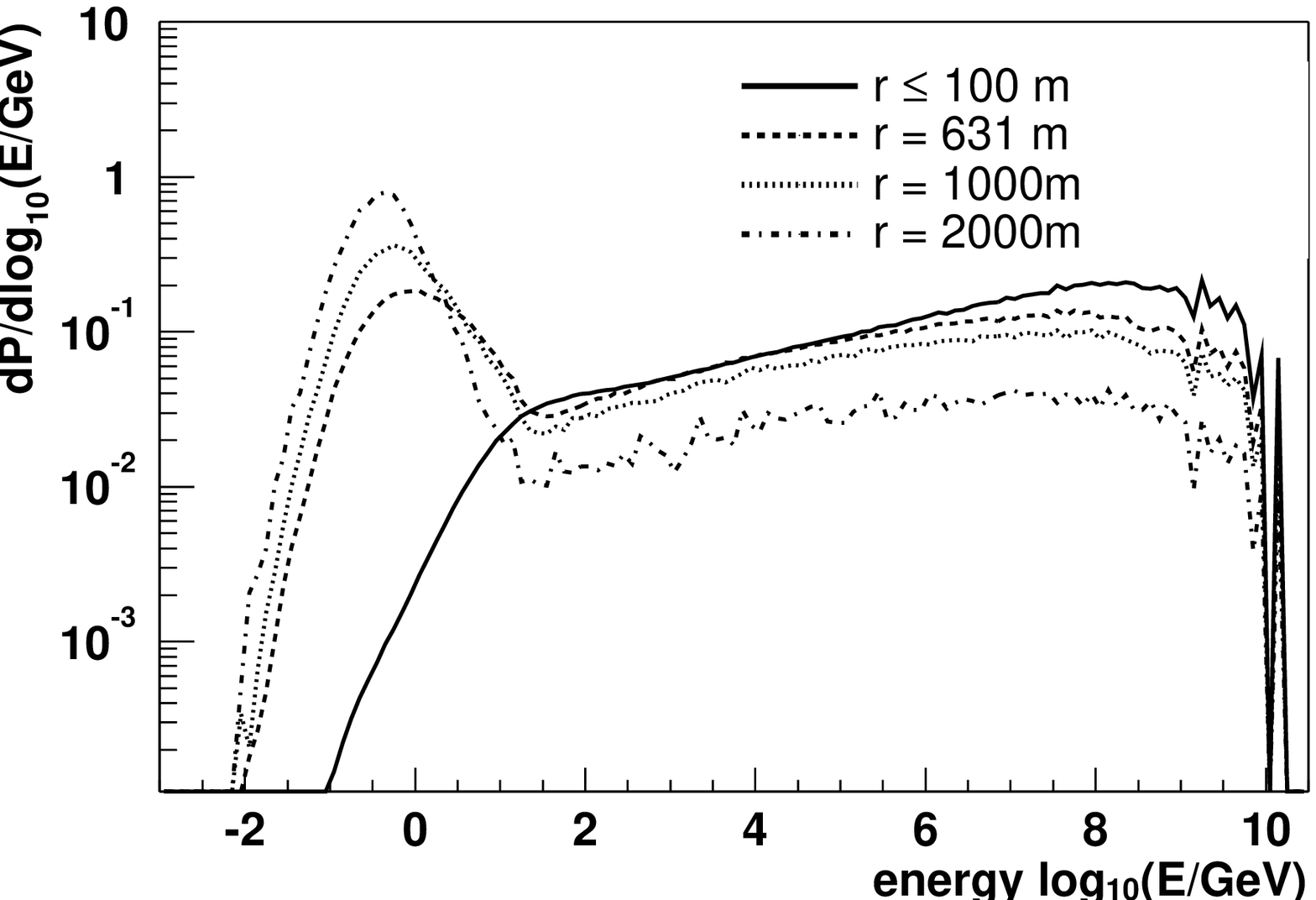}} \par}

{\centering \resizebox*{10cm}{!}{\includegraphics{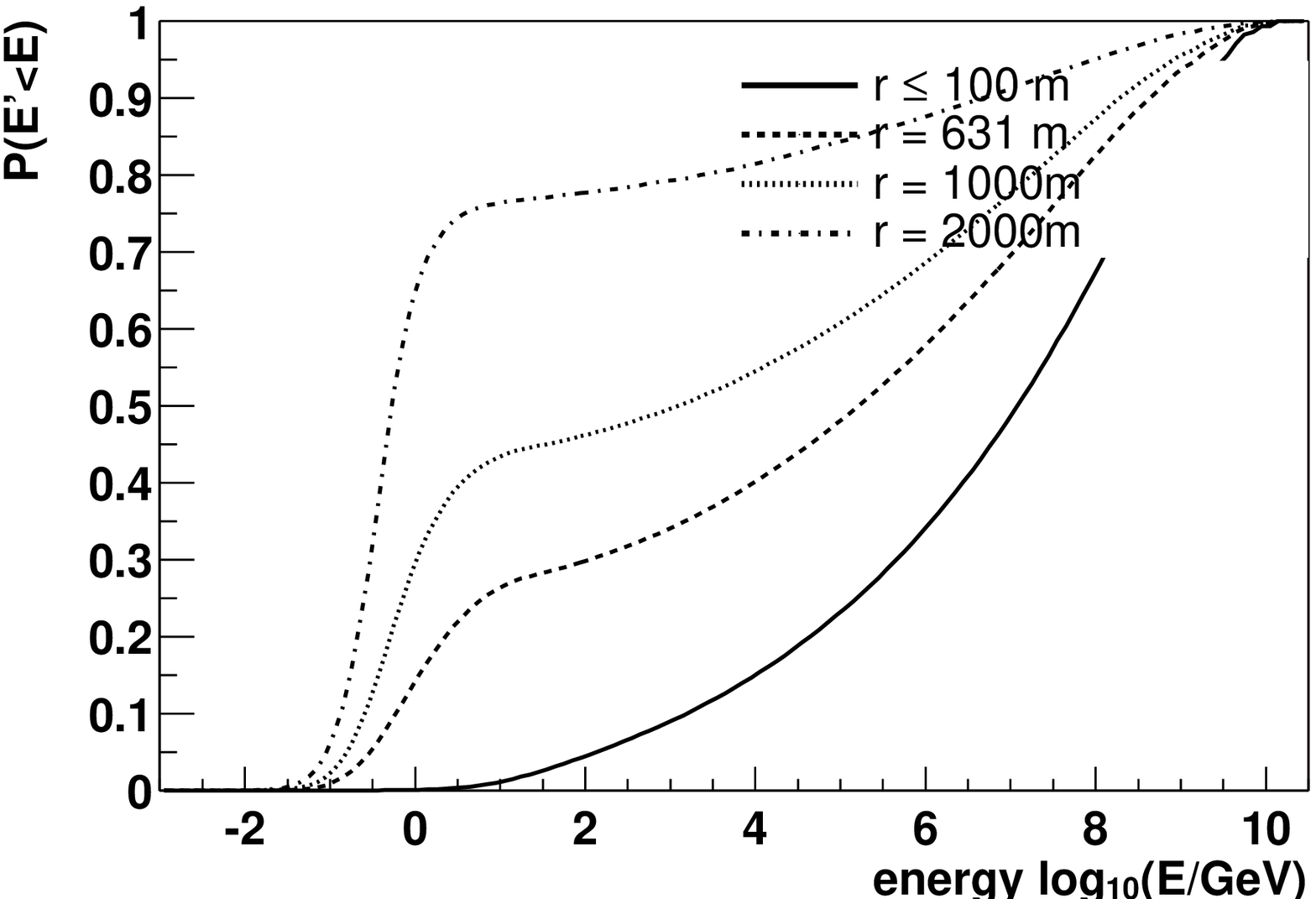}} \par}

\caption{\label{fig:ne_engy}The top figure shows the probability distribution
of initiators for ground level electrons as a function of the initiator's
energy for different radial distances from the core. The bottom figure
shows the integrated probability \protect\( P(E'<E)\equiv \int _{0}^{E}\frac{dP}{dE'}dE'\protect \)
which corresponds to the fraction of initiators with energy less than
\protect\( E\protect \). At r=2000m 65\% of the electron density
comes from low energy gammas with \protect\( E'<1\protect \) GeV.}
\end{figure}

\begin{figure}
{\centering \resizebox*{10cm}{!}{\includegraphics{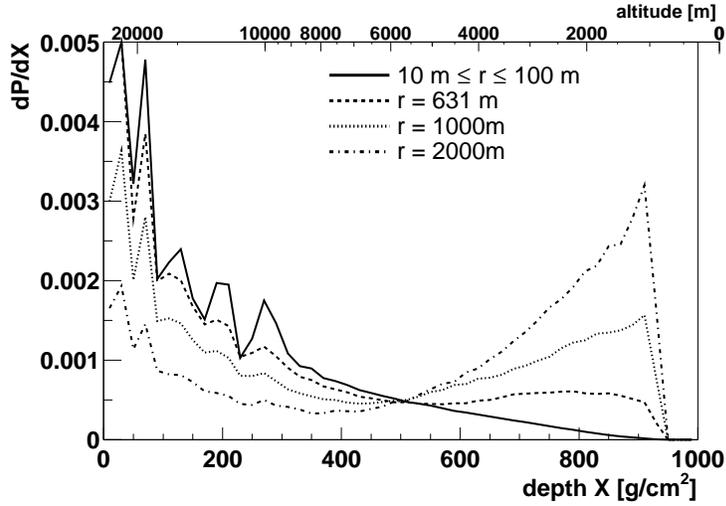}} \par}

\caption{\label{fig:ne_depth} The depth of electron initiators for 4 different
distances from the shower core. The greater is the distance from the
core, the larger the contribution from low altitudes. }
\end{figure}

\begin{figure}
{\centering \resizebox*{10cm}{!}{\includegraphics{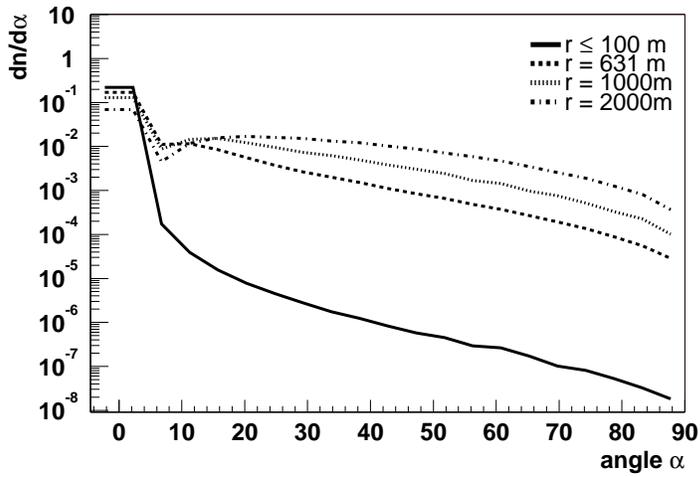}} \par}

\caption{\label{fig:angle} The emission angle of electron initiators as angle
with respect to the main shower axis. The first bin has been duplicated
in negative values for better visibility. Close to the core, the initiators
are predominantly emitted collinear with the shower axis, while at
larger distances greater emission angles are dominant. }
\end{figure}

\begin{figure}
{\centering \resizebox*{10cm}{!}{\includegraphics{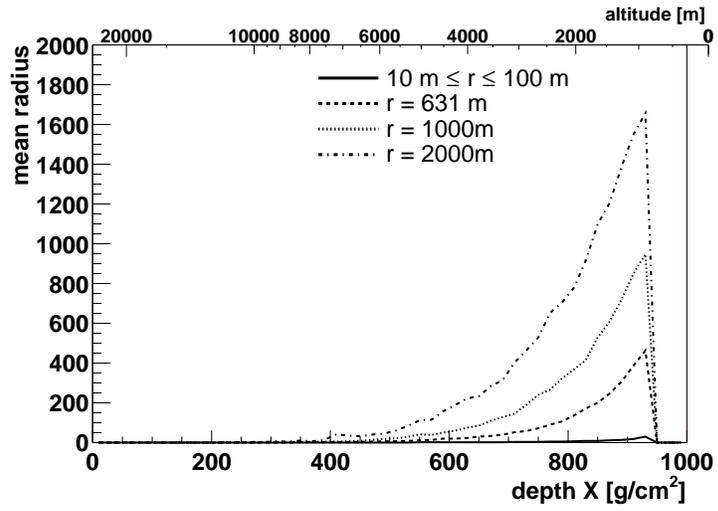}} \par}

\caption{\label{fig:1_mean_radius} The mean distance of electron initiators
to the shower axis. }
\end{figure}

We can analyze this interesting behavior in greater detail by looking
at the probability distributions as a function of energy and depth
separately. Figure \ref{fig:ne_engy} shows the distribution of energies
of electron initiators whose electromagnetic sub-showers contribute
to the electron density at 4 different lateral distances. The area
under each curve is normalized to 1 to get probability distributions.
Close to the shower core the most important contributions come from
high energies. At larger distances the contribution of low energy
initiators becomes more and more important. Fig. \ref{fig:ne_depth}
gives the probability distribution as a function of the altitude of
the initiators. At large distances from the core the dominant contribution
comes from relatively low altitudes, i.e. large depth. This confirms
that the power of lateral spreading for each high energy electromagnetic
cascade is limited. In fact, the greatest contribution at large lateral
distance comes from low energy sub-showers which are emitted at a
large angle with respect to the main shower axis, as seen in Fig.
\ref{fig:angle} and at a large distance with respect to the shower
axis (Fig. \ref{fig:1_mean_radius}). It is useful to read Figs. \ref{fig:ne_depth}
and \ref{fig:1_mean_radius} together, to see at which altitudes the
initiators are produced, and then to read off the mean distance from
the shower axis. One realizes that high altitude initiators are produced
close to the axis, and low altitude ones close to the point where
the electrons reach the ground. This means that the low energy hadronic
part of the shower is producing a significant portion of the lateral
expansion. 

Fig. \ref{fig:ne_depth} shows a very irregular structure below \( 300g/cm^{2} \).
This is a statistical fluctuation caused by some events having a very
energetic photon, which is consequently a sub-shower initiator for
many electrons and photons at ground level.

We note that the results on electron initiators presented here suggest
why analytic treatments of the lateral distribution function of the
air showers of hadronic primaries using a formula like the NKG \cite{NKG1,NKG2}
are unsatisfactory, even though the formula may perfectly describe
an electromagnetic sub-shower. The implementation of such analytic
approaches generally assumes that each sub-shower initiator has a
small angle with respect to the shower axis. As we have seen, this
is not correct when describing electrons and positrons at large distances
from the shower core. Only an implementation of NKG with full consideration
of the emission angle of each sub-shower initiator, without small
angle approximation, and taking the correct intersection of the inclined
sub-shower with the ground level, can be expected to give reasonable
results.

\subsection{Ground photons}

\begin{figure}
{\centering \resizebox*{10cm}{!}{\includegraphics{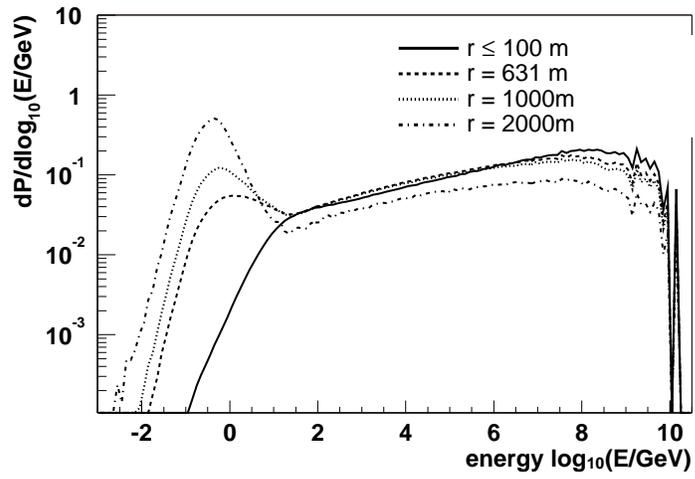}} \par}

\caption{\label{fig:np_engy}Energy distribution of photon initiators.}
\end{figure}

\begin{figure}
{\centering \resizebox*{10cm}{!}{\includegraphics{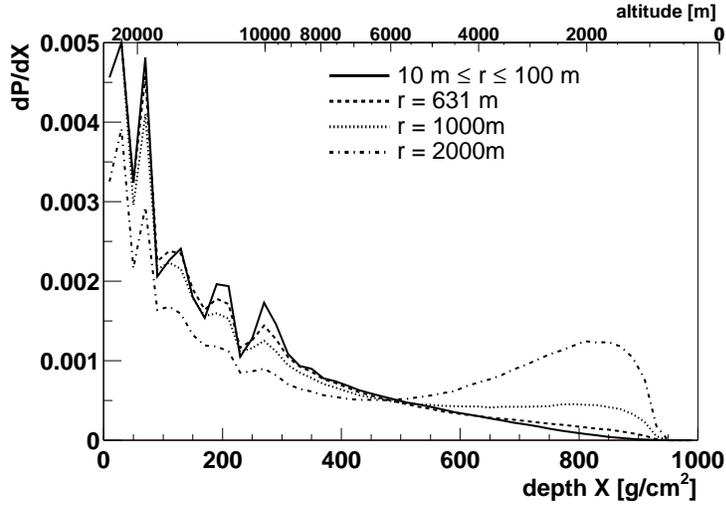}} \par}

\caption{\label{fig:np_depth} The distribution of vertical depth of photon
initiators.}
\end{figure}

Photons show similar behavior to electrons and positrons, since they
are created in the same electromagnetic sub-shower. However, Figs.
\ref{fig:np_engy} and \ref{fig:np_depth} reveal that the effect
discussed above for electrons is much less pronounced. Photon initiators
are produced higher in the atmosphere than electron initiators, even
though they are part of the same electromagnetic sub-showers. This
is probably due to the fact that photons do not lose energy by ionization
loss, therefore low energy photons have a higher chance to propagate
a large distance than electrons do.

\subsection{Ground muons}

\begin{figure}
{\centering \resizebox*{10cm}{!}{\includegraphics{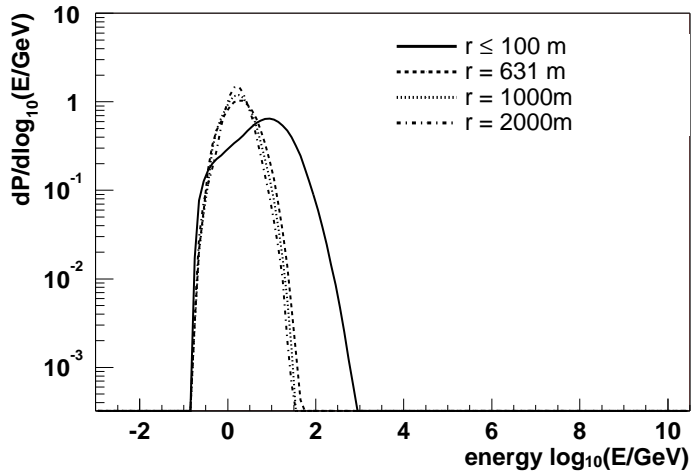}} \par}

\caption{\label{fig:nm_engy}This plot shows the energy distribution of muon
initiators, which is the energy of muons at the point of creation,
leading to ground level muons at different lateral distances.}
\end{figure}

\begin{figure}
{\centering \resizebox*{10cm}{!}{\includegraphics{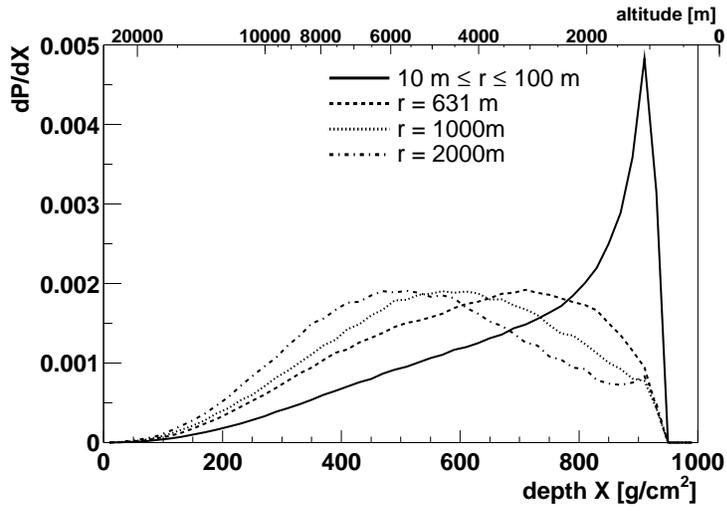}} \par}

\caption{\label{fig:nm_depth}The depth of muons at the point of creation.}
\end{figure}

The corresponding results for muons can be seen in Figs. \ref{fig:nm_engy}
and \ref{fig:nm_depth}. Their energy spectrum (Fig. \ref{fig:nm_engy})
at the point of creation gets peaked at a lower energy as one selects
muons arriving at larger distance from the shower axis, because a
lower energy makes larger emission angles more probable. As for the
depth of creation, one can see in Fig. \ref{fig:nm_depth} that at
high altitudes muon initiators contribute more at large distances
from the core, but never to the extent of electromagnetic initiators.
This is not surprising since the hadronic shower has to evolve down
in energy in order for pions or kaons to decay. 

\begin{figure}
{\centering \resizebox*{10cm}{!}{\includegraphics{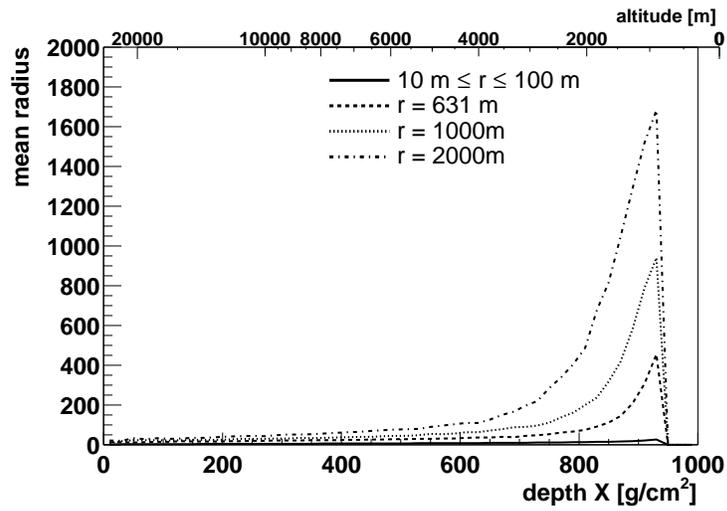}} \par}

\caption{\label{fig:2_mean_radius} The mean distance of muon initiators to
the shower axis. }
\end{figure}

A comparison of Figs. \ref{fig:nm_depth} and \ref{fig:2_mean_radius}
shows that muons observed at large radial distance are preferentially
but not exclusively produced close the the shower axis, propagating
to larger distances due to their large angle of production with respect
to the shower axis. 

The small peak in Fig. \ref{fig:nm_depth} at r=2000m and very low
altitudes may be a fluctuation or the onset of a low altitude contribution
as observed for electron densities; however the phenomenon is much
less pronounced and we do not investigate it further here.

\section{Hadronic reaction energy}

\begin{figure}
{\centering \resizebox*{10cm}{!}{\includegraphics{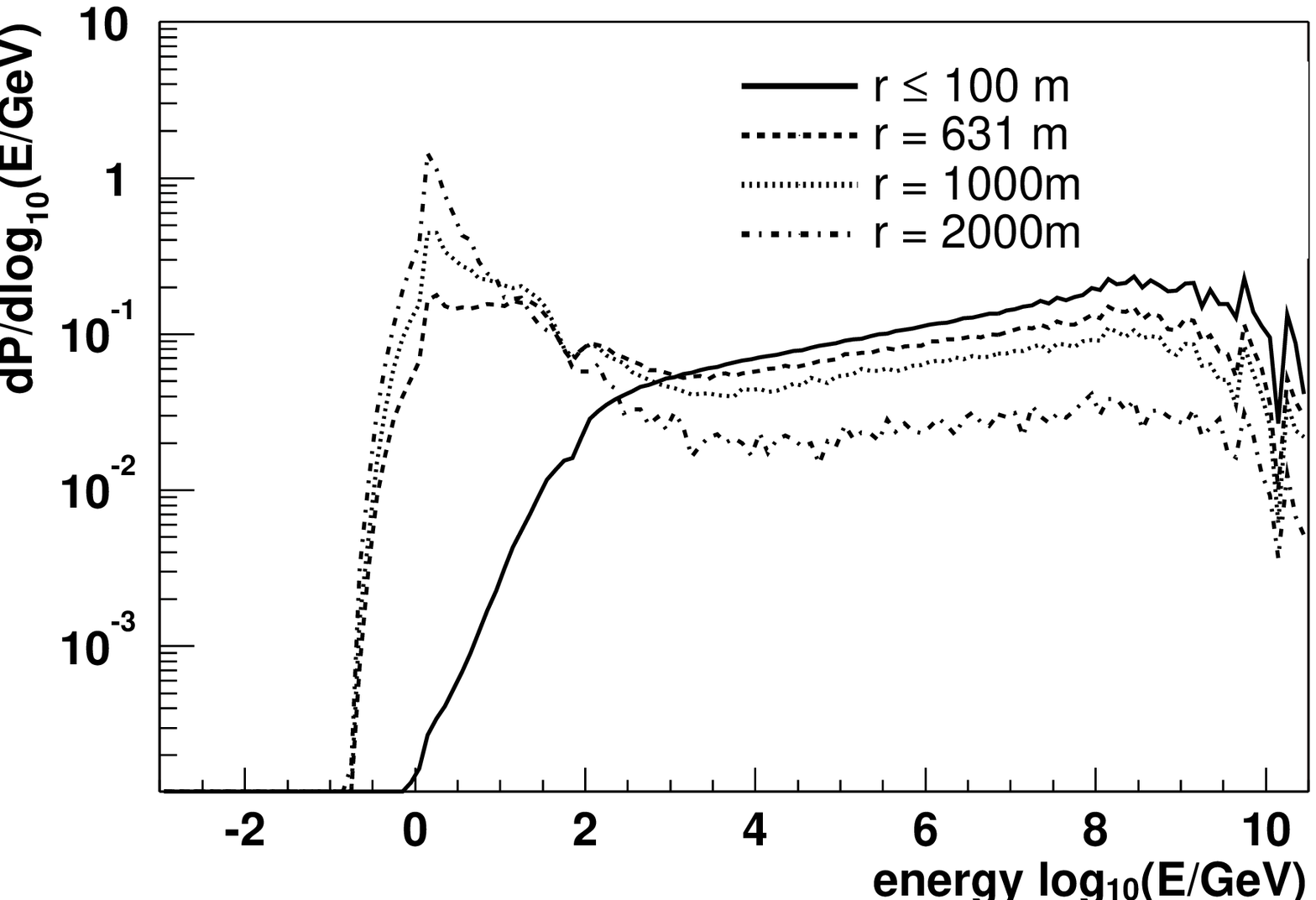}} \par}

{\centering \resizebox*{10cm}{!}{\includegraphics{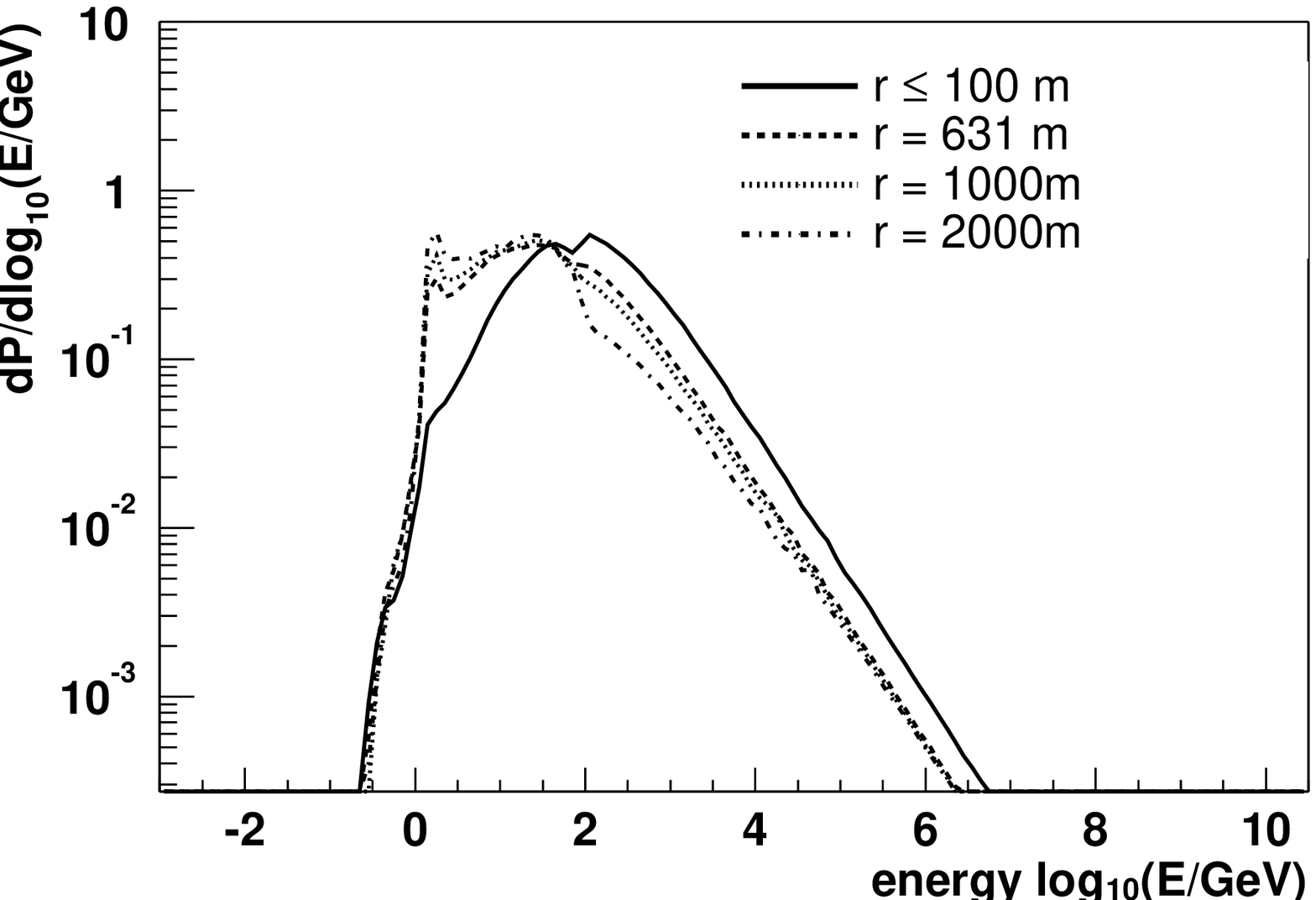}} \par}

\caption{\label{fig:reac_engy}Energy of the last hadronic reaction before
creating a electromagnetic (top) and a muonic (bottom) sub-shower.}
\end{figure}

It is also interesting to investigate the energy of the hadronic reactions
which are responsible for creating the sub-shower initiators. Figure
\ref{fig:reac_engy} shows the energy of the projectile in the lab
frame, of the hadron-air collision which creates the electron- or
muon-initiator. It reveals that high energy hadronic physics governs
the production of the electromagnetic component of the ground level
shower close to the core, but for larger distances from the core the
last hadronic reaction is in the low energy (E<100 GeV) regime. 

\begin{figure}
{\centering \resizebox*{10cm}{!}{\includegraphics{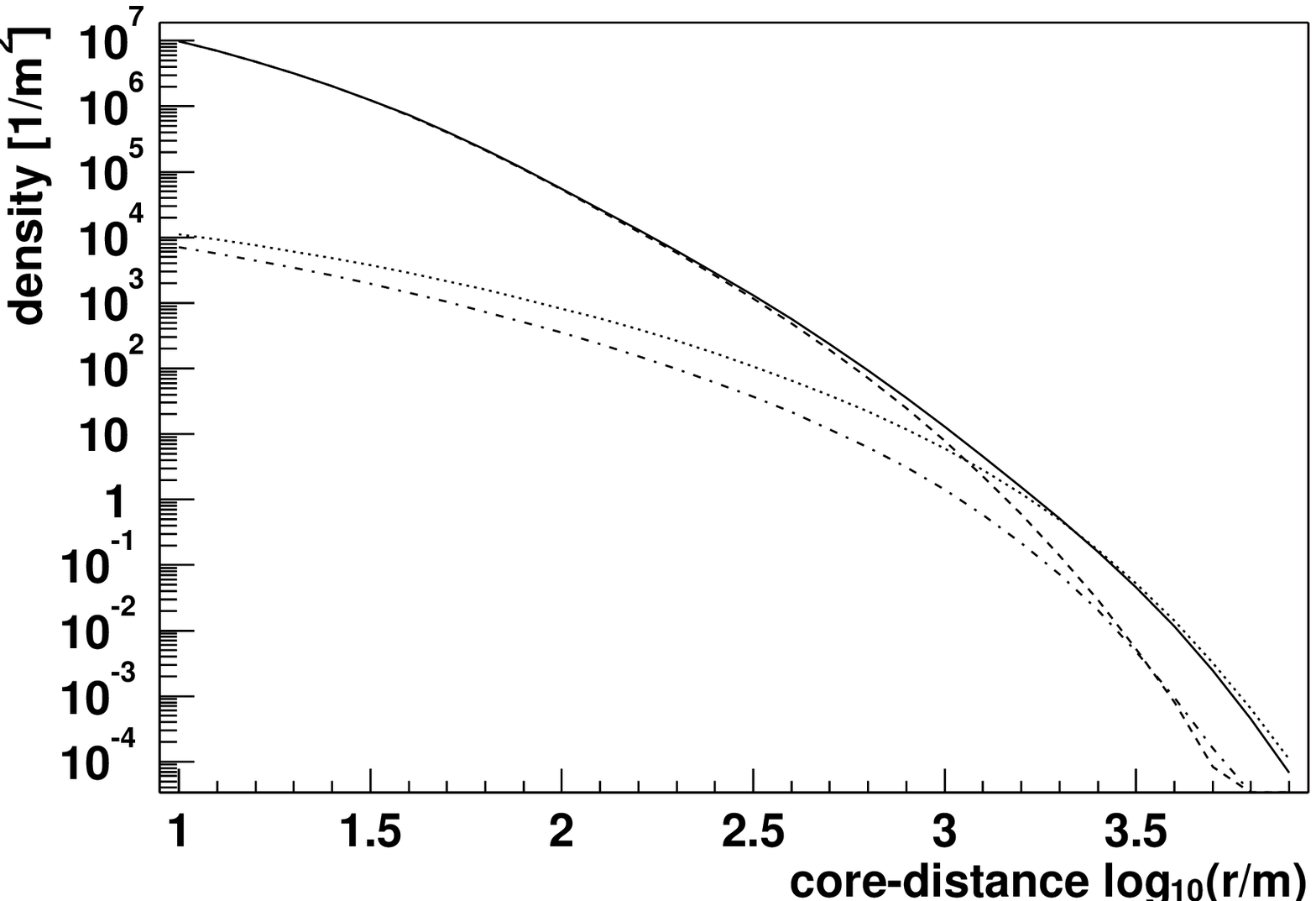}} \par}

{\centering \resizebox*{10cm}{!}{\includegraphics{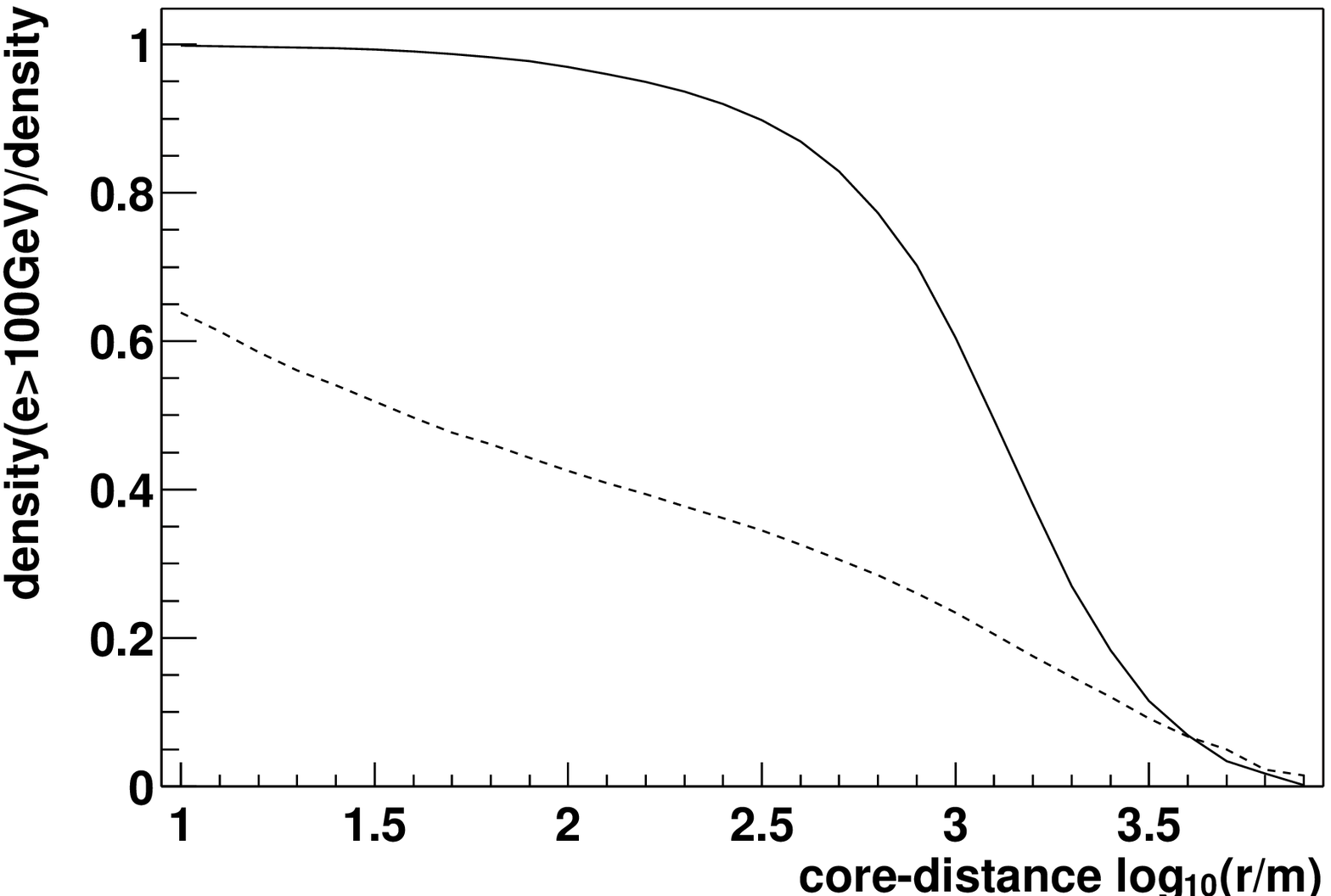}} \par}

\caption{\label{fig:ratio} The importance of high energy reactions to the
density of electrons and muons as a function of distance from the
shower core: the full line in the top figure shows the total density
of electrons as a function of distance from the shower core. The dashed
line represents the density of electrons whose initiators are created
in a high energy reaction (\protect\( E_{\mathrm{lab}}>100\protect \)
GeV) . The corresponding curves for muons are plotted as dotted and
dashed-dotted lines. The bottom figure shows the ratios for electrons
(full line) and muons (dashed line). }
\end{figure}

Figure \ref{fig:ratio} shows that at large distances from the core,
low energy reactions become increasingly important for determining
particle densities. The top figure shows the total lateral distribution
function of electrons plotted in a full line, while the part contributed
by initiators which are created in an hadronic reaction with an energy
\( E_{\mathrm{lab}}>100 \) GeV is shown with a dashed line. The corresponding
curves for muons are drawn as dotted (muon lateral distribution function)
and dashed-dotted lines (contribution by high energy reactions with
\( E_{\mathrm{lab}}>100 \) GeV only). The bottom figure shows the
ratios of the corresponding curves, for electrons and for muons. At
600m 20\% of the electron density comes from a low energy hadronic
reaction; at 1km this fraction is 40\%. The situation is even more
pronounced in the case of muons, due to the comparatively long lifetime
of charged mesons. Here the last hadronic reaction is in most cases
a low energy reaction, even at small distances from the core. 

These results shed new light on the importance of low energy hadronic
modeling in air showers, an aspect which is often neglected in the
literature. The impact of low energy shower modeling can be directly
seen in Fig. \ref{fig:reac_engy}. In the simulation program used
for this study, we employed the low energy hadronic model GHEISHA
to describe interactions of projectiles whose lab energy is less than
90 GeV, while at higher energies QGSJET is used. This follows the
CORSIKA procedure, only differing by choosing the energy of transition
between low and high energy hadronic models at 90 GeV rather than
80 GeV. Fig. \ref{fig:reac_engy} shows that the projectile energy
distributions show a break at \( \log _{10}(E/\mathrm{GeV})\approx 2 \)
, especially for muons (lower figure) at large distance from the core.
The sharp break is an artifact of a discontinuity in the modeling
of the hadronic interactions, between QGSJET and the low energy hadronic
model GHEISHA. Evidently, at the transition energy \( E_{\mathrm{lab}}=90 \)
GeV, GHEISHA produces many more muons at large distances than QGSJET;
this arises because the two models have different differential cross
sections and particle production levels. (For examples of differences
between these models in this energy region, see \cite{sciutto}.)
With correct modeling, the high and low energy models must match at
the transition energy, in principle in every detail. What we see here
is that at large radius, these details affect the predicted shower
properties. We have checked, however, that the qualitative conclusion
that low energy, low altitude initiators dominate for large radius
electrons, is not an artifact of the non-smooth transition between
low and high energy hadronic models. 

It has already been pointed out that for much lower energy primaries
the low energy hadronic model is important for muons measured at ground
level. In Ref. \cite{Engel01,Engel99} it is shown that for \( 10^{15} \)eV
hadronic showers, the energy of the last hadronic reaction before
producing a muon (\( E_{\mu }>1 \) GeV) is in 50\% of all cases a
low energy reaction with \( E_{\mathrm{lab}}<100 \) GeV. Refs. \cite{Haungs01,Swordy:2002df}
study the low energy hadronic model dependence by analyzing CORSIKA
\cite{CORSIKA} simulations of \( 10^{15} \) eV hadron induced showers
with two different models at energies below 80 GeV: GHEISHA\cite{GHEISHA}
and UrQMD \cite{Bass:1998ca,Bleicher:1999xi}. The two models show
differences in the energy spectrum of produced muons, but the overall
number of muons stays approximatively the same in both cases \cite{Swordy:2002df}.
This fact becomes important when measuring muons with different energy
thresholds as shown in Ref. \cite{Haungs01}. While possibly not surprising,
the analysis presented here shows that the low energy hadronic model
is also important for determining the lateral distribution function
at large radius for ultrahigh energy cosmic ray showers.

\section{Mean arrival time of ground particles}

We have seen that in the case of lateral distribution functions of
electrons, there are two distinct regions in altitude and energy which
contribute differently as a function of distance from the shower axis.
This gives rise to the question where the time delay of particles
with respect to the shower front comes from. In a simplified model
one can consider three contributions: The time delay is caused by

\begin{figure}
{\centering \resizebox*{5cm}{!}{\includegraphics{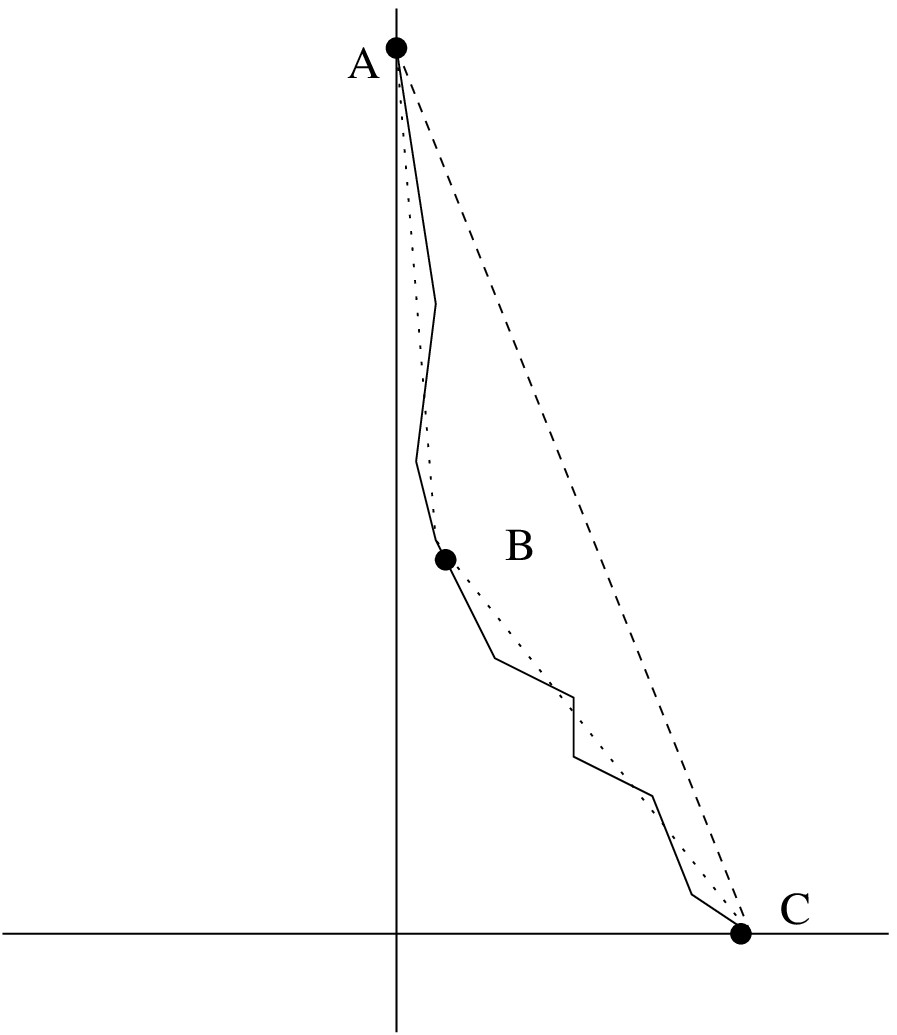}} \par}

\caption{\label{fig:timedelay}The contributions to the time delay of particles:
hadronic time delay from A to B, electromagnetic/muonic from B to
C. The difference of the sum (dotted line) to the total time delay
(dashed line) represents the geometric delay as explained in the text. }
\end{figure}

\begin{itemize}
\item hadronic showering,
\item electromagnetic showering, and
\item geometry.
\end{itemize}
Fig. \ref{fig:timedelay} shows illustrates the different parts. The
hadronic time delay is defined as \( \Delta t_{\mathrm{had}}=t_{AB}-\overline{AB}/c \)
with \( \overline{AB} \) being the distance from A to B and \( t_{AB} \)
being the actual time spent in the hadronic cascade prior to production
of the initiator, i.e. going from point A (first interaction) to point
B (production of the initiator, e.g. a photon or a muon). \( \Delta t_{\mathrm{em}}=t_{BC}-\overline{BC}/c \)
is the corresponding time delay of the electromagnetic (or muonic)
sub-shower. The total time delay is defined as \( \Delta t_{\mathrm{tot}}=t_{AB}+t_{BC}-\overline{AC}/c \).
And finally the geometric time delay is \( \Delta t_{\mathrm{geo}}=(\overline{AB}+\overline{BC}-\overline{AC})/c=\Delta t_{\mathrm{tot}}-\Delta t_{\mathrm{had}}-\Delta t_{\mathrm{em}} \)
. A detailed Monte Carlo model permits keeping track of each propagating
particle and hence computing the values \( \Delta t_{\mathrm{tot}},\Delta t_{\mathrm{had}},\Delta t_{\mathrm{em}} \).

\begin{figure}
{\centering \resizebox*{8cm}{!}{\includegraphics{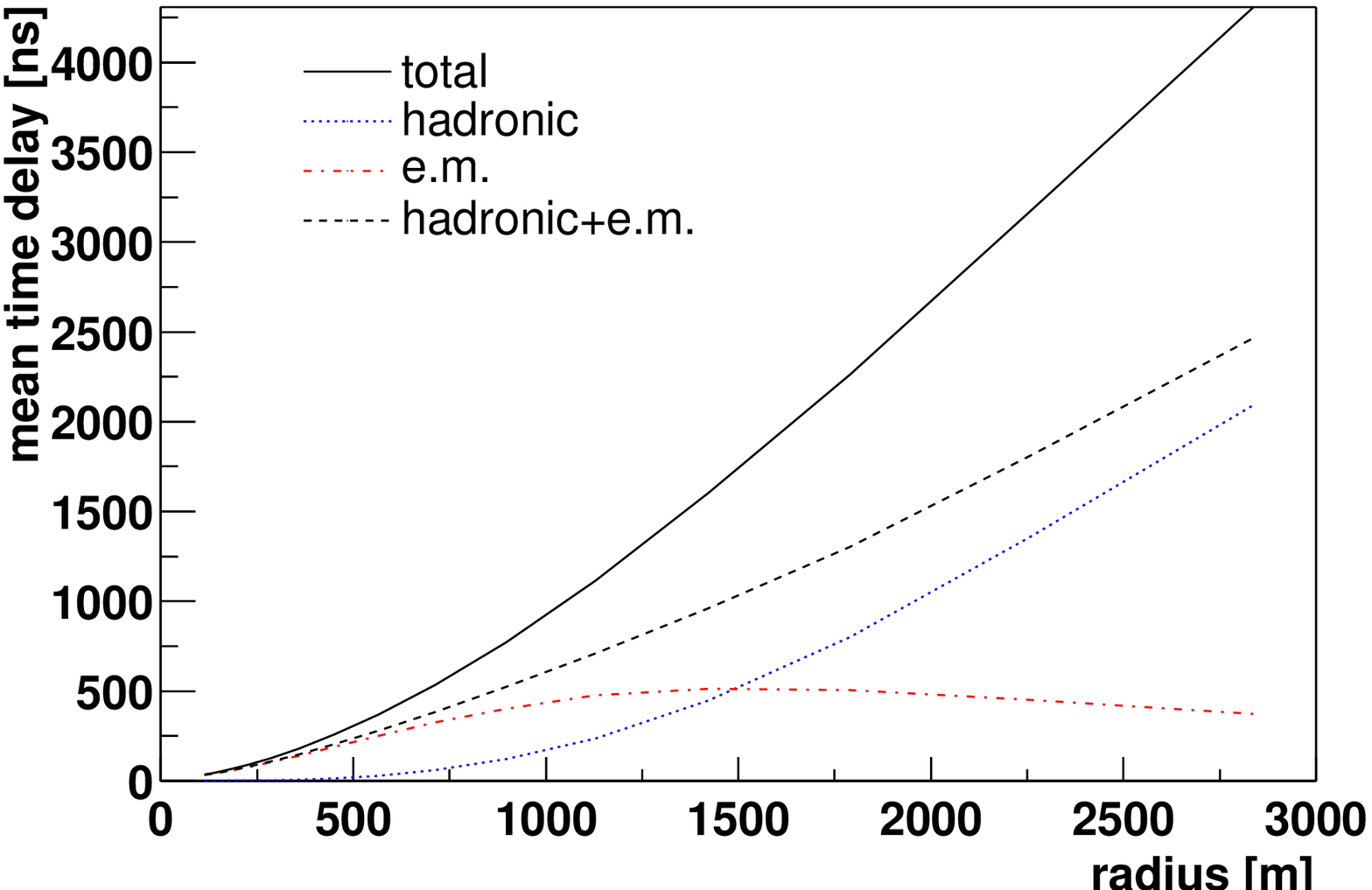}} \par}

{\centering \resizebox*{8cm}{!}{\includegraphics{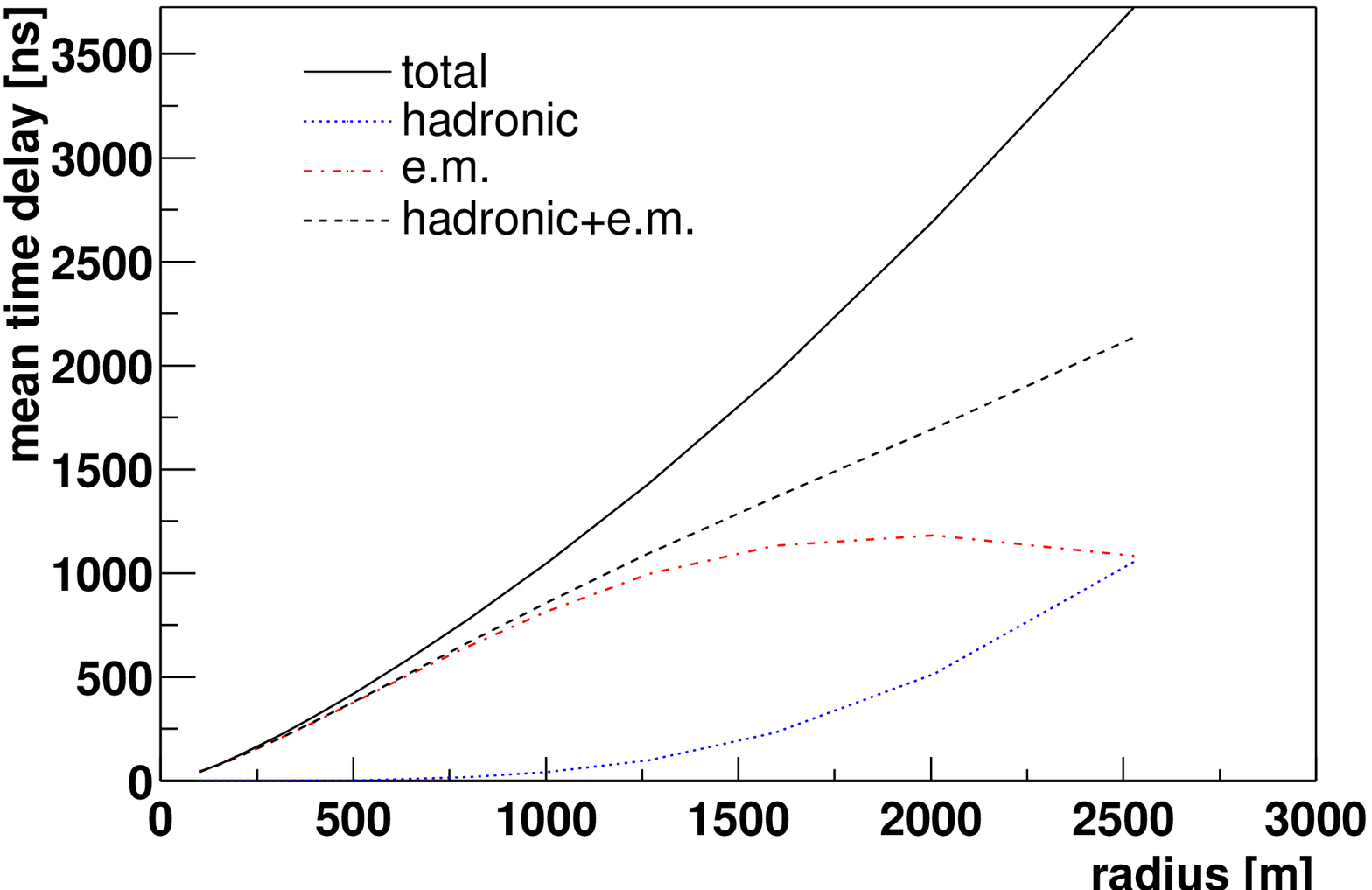}} \par}

{\centering \resizebox*{8cm}{!}{\includegraphics{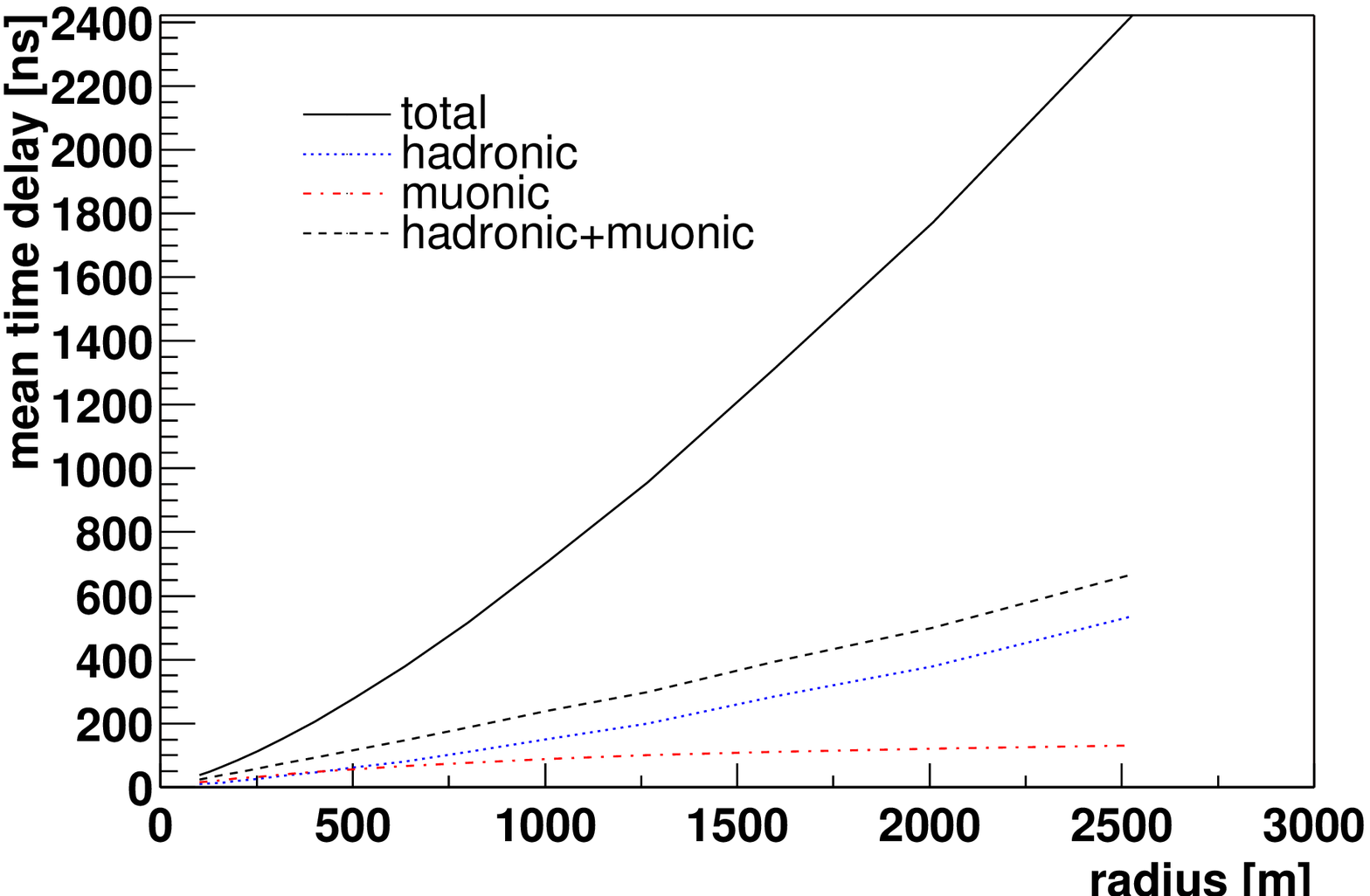}} \par}

\caption{Contributions to the mean time delay of electrons (top), photons
(middle) and muons (bottom) as a function of the distance from the
shower axis. \label{fig:timedelay2}}
\end{figure}

Fig. \ref{fig:timedelay2} shows the results, which are best understood
by referencing also the figures of the previous sections giving production
height and radius of the initiators. Ground electrons close to the
core spend most of their time in the electromagnetic sub-shower, since
for them the main contribution comes from high altitude high energy
initiators; the hadronic part of their history is at high energy for
which multiple collisions do not increase the path length significantly,
so their hadronic time delay is small. Further away from the core
the hadronic part of the time delay increases, becoming as important
as the electromagnetic one at 1500m. The increasing hadronic time
delay is correlated with an increasing time delay caused by geometry,
since both the radius of production of the initiators and their average
emission angle increases with radius of the observed electron. Thus
at very large distances, geometry and hadronic time delay are the
dominant contributions, and the electromagnetic spreading is roughly
constant as we have seen earlier.

As for ground photons, the main time delay is caused by the electromagnetic
sub-shower. The difference from electrons becomes clear when one remembers
that photon initiators are higher in the atmosphere than electron
initiators (see Figs. \ref{fig:ne_depth} and \ref{fig:np_depth}).

The time delay of muons is mainly caused by geometry and, though less
important, the hadronic part of the shower.

\section{Conclusions}

We have analyzed in detail how electromagnetic and muonic sub-showers
contribute to the overall lateral distribution functions of an ultrahigh
energy cosmic ray air shower. At all distances from the core, ground
level muons come mostly from low energy muons emitted at intermediate-high
altitudes in the atmosphere. However, at large distances from the
core (>1000m), electrons come predominantly from low energy \( \pi ^{0}/\eta  \)
's produced at low altitudes and large distance from the core, and
with a large angle with respect to the shower axis. Moreover the collisions
producing the initiators are themselves low energy hadron-nucleus
interactions which in standard air shower simulations are treated
by a different model than the high energy interactions. 

We conclude that in order to achieve detailed agreement between data
and models for the lateral distribution functions and arrival-delay
time distributions, necessary in order to optimally reconstruct properties
of the primary, it is necessary to take greater care for modeling
the low energy hadronic interactions in the showers.

\subsection*{Acknowledgments}

This research was supported by NASA grant NAG-9246, NSF-PHY-9996173,
and NSF-PHY-0101738. The computations were made on NYU's Mafalda:
a Linux cluster financed in part by the Major Research Instrumentation
grant NSF-PHY-0116590.

We would like to thank A.A. Watson and R. Engel for useful comments
on the first draft, as well as D.Heck for further informations.

\bibliographystyle{unsrt}
\bibliography{lat}

\begin{thebibliography}{10}

\bibitem{QSJET}
N.N. Kalmykov, S.S. Ostapchenko, and A.I. Pavlov.
\newblock {\em Nucl. Phys. B}, 17, 1997.

\bibitem{GHEISHA}
H.Fesefeldt.
\newblock The simulation of hadronic showers.
\newblock {\em PITHA 85/02}, Aachen 1985.

\bibitem{EGS4}
W.~R. Nelson, H.~Hirayama, and D.W.O.
\newblock {\em the EGS4 Code System}.
\newblock SLAC-265, Stanford Linear Accelerator Center, 1985.

\bibitem{CORSIKA}
D.~Heck, J.~Knapp, J.N. Capdevielle, G.~Schatz, and T.~Thouw.
\newblock {\em Report FZKA 6019}, 1998.

\bibitem{NKG1}
K.~Kamata and J.~Nishimura.
\newblock {\em Progr. Theor. Phys.}, Suppl. 6, 1958.

\bibitem{NKG2}
K.~Greisen.
\newblock {\em Prog. Cosmic Ray Physics}, 3, 1956.

\bibitem{sciutto}
S.J. Sciutto.
\newblock The {Aires} system for shower simulations.
\newblock In {\em Prceedings of the 27th International Comsmic Ray Conference}.

\bibitem{Engel01}
R.~Engel.
\newblock rapporteur talk: Particle and interaction physics.
\newblock In {\em proceedings of the 27th International Cosmic Ray Conference,
  Hamburg, Aug. 7-15, 2001}.

\bibitem{Engel99}
R.~Engel, T.K. Gaisser, and T.~Stanev.
\newblock Extensive air showers and hadronic interaction models.
\newblock In I.~Sarcevic and C.-I. Tan, editors, {\em proceedings of ISMD99 --
  International Symposium on Multiparticle Dynamics, Providence, Rhode Island,
  August 9-13, 1999}. World Scientific.

\bibitem{Haungs01}
A.~Haungs et~al.
\newblock The primary energy spectrum of cosmic rays obtained by muon density
  measurement at kascade.
\newblock In {\em proceedings of the 27th International Cosmic Ray Conference,
  Hamburg, Aug. 7-15, 2001}.

\bibitem{Swordy:2002df}
S.~P. Swordy et~al.
\newblock The composition of cosmic rays at the knee.
\newblock {\em astro-ph/0202159}, 2002.

\bibitem{Bass:1998ca}
S.~A. Bass et~al.
\newblock Microscopic models for ultrarelativistic heavy ion collisions.
\newblock {\em Prog. Part. Nucl. Phys.}, 41:225--370, 1998.

\bibitem{Bleicher:1999xi}
M.~Bleicher et~al.
\newblock Relativistic hadron hadron collisions in the ultra- relativistic
  quantum molecular dynamics model.
\newblock {\em J. Phys.}, G25:1859--1896, 1999.

\end{thebibliography}

\end{document}